\documentclass[10pt,onecolumn,twoside]{article}

\usepackage{amsmath}
\usepackage{graphicx}
\usepackage{amssymb}
\usepackage{bm}
\usepackage{mathtools}
\usepackage{cite}
\usepackage{amsthm}
\usepackage[hidelinks]{hyperref}
\usepackage{float}
\usepackage{fullpage}
\usepackage[font=small,labelformat=empty]{caption}

\def\defn{\,\coloneqq\,}

\def\argmin{\mathop{\mathrm{arg\,min}}} 

\def\Im{{\mathsf{Im}}}
\def\prox{\mathsf{prox}}
\def\lim{\mathop{\mathsf{lim}}}
\def\log{\mathsf{log}}

\def\df{g} 
\def\h{{h_{\textsf{\tiny mmse}}}} 

\def\E{\mathbb{E}}
\def\R{\mathbb{R}}

\def\xbm{{\bm{x}}}
\def\zbm{{\bm{z}}}
\def\ybm{{\bm{y}}}

\def\ebm{{\bm{e}}}

\def\ubm{{\bm{u}}}

\def\zerobm{{\bm{0}}}

\def\sbm{{\bm{s}}}

\def\nbm{{\bm{n}}}

\def\d{{\, \mathsf{d}}}

\def\Tsf{{\mathsf{T}}}
\def\Dsf{{\mathsf{D}}}
\def\Hsf{{\mathsf{H}}}
\def\Jsf{{\mathsf{J}}}

\def\dsf{{\mathsf{d}}}

\def\argmin{\mathop{\mathsf{arg\,min}}} 

\def\xbmhat{{\widehat{\bm{x}}}}

\def\Hbf{{\mathbf{H}}}

\def\Ibf{{\mathbf{I}}}

\def\Ncal{\mathcal{N}}

\def\Xcal{\mathcal{X}}

\newtheoremstyle{theoremdd}
{\topsep}
{\topsep}
{}
{0pt}
{\em}
{: }
{ }
{\thmname{#1}\thmnumber{ #2}\thmnote{ (#3)}}
\theoremstyle{theoremdd}

\theoremstyle{plain} 

\newtheorem{theorem}{Theorem}

\newtheorem{assumption}{Assumption}

\title{Provable Convergence of Plug-and-Play Priors\\with MMSE denoisers} 

\author{Xiaojian~Xu%
\thanks{Department of Computer Science \& Engineering, Washington University in St.~Louis, St.~Louis, MO 63130.}
\hspace{0.05em},
Yu~Sun$^\ast$,
Jiaming~Liu%
\thanks{Department of Electrical \& Systems Engineering, Washington University in St.~Louis, St.~Louis, MO 63130.}
\hspace{0.05em}, 
Brendt~Wohlberg%
\thanks{Theoretical Division, Los Alamos National Laboratory, Los Alamos, NM 87545 USA.}
\hspace{0.05em}, 
and Ulugbek~S.~Kamilov$^{\ast, \dagger}$}

\begin{document}
\date{}
\maketitle


\begin{abstract}
Plug-and-play priors (PnP) is a methodology for regularized image reconstruction that specifies the prior through an image denoiser. While PnP algorithms are well understood for denoisers performing \emph{maximum a posteriori probability (MAP)} estimation, they have not been analyzed for the \emph{minimum mean squared error (MMSE)} denoisers. This letter addresses this gap by establishing the first theoretical convergence result for the iterative shrinkage/thresholding algorithm (ISTA) variant of PnP for MMSE denoisers. We show that the iterates produced by PnP-ISTA with an MMSE denoiser converge to a stationary point of some global cost function. We validate our analysis on sparse signal recovery in compressive sensing by comparing two types of denoisers, namely the \emph{exact} MMSE denoiser and the \emph{approximate} MMSE denoiser obtained by training a deep neural net.
\end{abstract}


\section{Introduction}
The recovery of an unknown signal from its noisy measurements is fundamental in signal processing. It often arises in the context of linear inverse problems, where the goal is to recover $\xbm \in \R^n$ from its noisy measurements 
\begin{equation}
\label{Eq:inverse}
\ybm = \Hbf \xbm + \ebm\;.
\end{equation}
Here, $\Hbf\ \in\R^{m \times n}$ represents the response of the acquisition system and $\ebm \in \R^n$ models the measurements noise.

The solution of ill-posed inverse problems is commonly formulated as a \emph{regularized inversion}, expressed as an optimization problem
\begin{equation}
\label{Eq:Optimization}
\xbmhat = \argmin_{\xbm \in \R^n} f(\xbm) \quad\text{with}\quad f(\xbm) = \df(\xbm)
+ h(\xbm) \;,
\end{equation}
where $\df$ is the data-fidelity term and $h$ is the regularizer or prior. Proximal algorithms~\cite{Parikh.Boyd2014} are widely used for solving~\eqref{Eq:Optimization} when the  regularizer is nonsmooth. For example, the \emph{iterative shrinkage/thresholding algorithm (ISTA)}~\cite{Figueiredo.Nowak2003, Bect.etal2004, Daubechies.etal2004, Beck.Teboulle2009a} is a standard approach for solving~\eqref{Eq:Optimization} via the iterations
\begin{subequations}
        \label{alg:ista}
        \begin{align}
        \zbm^t &= \xbm^{t-1}-\gamma \nabla \df(\xbm^{t-1})\\
        \xbm^t &= \prox_{\gamma h}(\zbm^t) \;,
        \end{align}
\end{subequations}
where $\gamma > 0$ is the step-size parameter. The second step of ISTA relies on the \emph{proximal operator} defined as
\begin{equation}
\label{Eq:ProximalOperator}
\prox_{\tau h}(\zbm) \defn \argmin_{\xbm \in \R^n}\left\{\frac{1}{2}\|\xbm-\zbm\|_2^2 + \tau h(\xbm)\right\},
\end{equation}
where $\tau > 0$ controls the influence of $h$. The proximal operator can be interpreted as a \emph{maximum a posteriori
probability (MAP)} estimator for the AWGN denoising problem
\begin{equation}
\label{Eq:xhat}
\zbm = \xbm + \nbm \quad\text{where}\quad \xbm \sim p_\xbm,\quad \nbm \sim \Ncal(\zerobm, \tau \Ibf) \;,
\end{equation}
by setting $h(\xbm) = -\log(p_\xbm(\xbm))$. This perspective inspired the development of the PnP methodology~\cite{Venkatakrishnan.etal2013, Sreehari.etal2016}, where the proximal operator is replaced by a more general denoiser $\Dsf(\cdot)$, such as BM3D~\cite{Dabov.etal2007} or DnCNN~\cite{Zhang.etal2017}. For example, PnP-ISTA~\cite{Kamilov.etal2017} can be summarized as
\begin{subequations}
        \label{alg:pnpista}
        \begin{align}
        \zbm^t &= \xbm^{t-1}-\gamma \nabla \df(\xbm^{t-1})\\
        \xbm^t &= \Dsf_\sigma(\zbm^t) \;,
        \end{align}
\end{subequations}
where by analogy to $\tau$ in~\eqref{Eq:ProximalOperator}, we introduce the parameter $\sigma > 0$ for controling the relative strength of the denoiser $\Dsf_\sigma$.

PnP algorithms have been shown to achieve state-of-the-art performance in many imaging problems.~\cite{Chan.etal2017, Ono2017, Zhang.etal2017a, Meinhardt.etal2017, Buzzard.etal2018, Teodoro.etal2019, Dong.etal2019, Chan2019, Sun.etal2019c, Song.etal2020}. Recent work has also provided theoretical convergence guarantees for PnP algorithms under various assumptions on the data-fidelity term and the denoiser~\cite{Chan.etal2017, Sreehari.etal2016, Buzzard.etal2018, Teodoro.etal2017, Sun.etal2019a, Ryu.etal2019, Gavaskar.Chaudhury2020}. However, PnP has \emph{not} been investigated for denoisers performing \emph{minimum mean squared error (MMSE)} estimation
\begin{equation}
\label{Eq:MMSEDenoiser}
\Dsf_\sigma(\zbm) = \E[\xbm | \zbm] = \int_{\R^n} \xbm p_{\xbm|\zbm}(\xbm|\zbm) \dsf \xbm \;.
\end{equation}
MMSE denoisers are ``optimal'' with respect to widely used image-quality metrics such as signal-to-noise ratio (SNR). However, they are generally \emph{not} nonexpansive~\cite{Teodoro.etal2019} and their direct computation is often intractable in high-dimensions~\cite{Kazerouni.etal2013}. Insights into the performance of PnP for MMSE denoisers are valuable as many denoisers (pre-trained CNNs, NLM, BM3D) can be interpreted as \emph{approximate} or \emph{empirical} MMSE denoisers~\cite{Buades.etal2010}. In this paper, we show that PnP-ISTA with an MMSE denoiser converges to a stationary point of a certain (possibly nonconvex) cost function. To the best of our knowledge, this explicit link between PnP-ISTA and MMSE estimation is absent from the current literature on PnP methods. Our analysis builds on an elegant formulation by Gribonval~\cite{Gribonval2011} that establishes a direct link between MMSE estimation and regularized inversion. We validate our analysis on sparse signal recovery in compressive sensing by comparing PnP-ISTA with two types of denoisers---the \emph{exact} MMSE denoiser and the \emph{approximate} MMSE denoiser obtained by training DnCNN~\cite{Zhang.etal2017} to minimize the mean squared error (MSE). Our simulations show convergence of PnP-ISTA for both denoisers, highlight close agreement between their performance, and illustrate the limitation of using an AWGN denoiser as a prior within ISTA.


\section{Theoretical Analysis}
\label{sec:results}

Our analysis requires three assumptions that serve as sufficient conditions for establishing theoretical convergence.
\begin{assumption}
\label{As:NonDeg}
The prior $p_\xbm$ is non-degenerate over $\R^n$.
\end{assumption}
\noindent
As a reminder, a probability distribution $p_\xbm$ is \emph{degenerate} over $\R^n$, if it is  supported on a space of lower dimensions than $n$. Consider the image set of the MMSE denoiser $\Xcal \defn \Im(\Dsf_\sigma)$. Assumption~\ref{As:NonDeg} is required for establishing an explicit link between~\eqref{Eq:MMSEDenoiser} and the following regularizer~\cite{Gribonval2011}
\begin{align}
\label{Eq:Regularizer}
&\h(\xbm) \defn\\
&\nonumber
\begin{cases}
-\frac{1}{2\gamma} \|\xbm-\Dsf_\sigma^{-1}(\xbm)\|^2 + \frac{\sigma^2}{\gamma}h_\sigma(\Dsf_\sigma^{-1}(\xbm)) & \text{for } \xbm \in \Xcal \\
+\infty &\text{for } \xbm \notin \Xcal \;,
\end{cases}
\end{align}
where $\gamma > 0$ is the step-size, $\Dsf_\sigma^{-1}: \Xcal \rightarrow \R^n$ is the inverse mapping, which is well defined and smooth over $\Xcal$ (see Appendix~\ref{Sec:MMSEDenoiser}), and $h_\sigma(\cdot) \defn -\log(p_\zbm(\cdot))$, where $p_\zbm$ is the probability distribution of the AWGN corrupted observation~\eqref{Eq:xhat}.
As discussed in Appendix~\ref{Sec:MMSEDenoiser}, the function $\h$ is smooth for any $\xbm \in \Xcal$, which is the consequence of the smoothness of both $\Dsf_\sigma^{-1}$ and $h_\sigma$.

\begin{assumption}
\label{As:Smooth}
The function $g$ is continuously differentiable and has a
Lipschitz continuous gradient with constant $L > 0$.
\end{assumption}
\noindent
This is a standard assumption used extensively in the analysis of gradient-based algorithms (see~\cite{Nesterov2004}, for example).

\begin{assumption}
\label{As:GlobMin}
The function $f$ has a finite infimum $f^\ast > -
\infty$.
\end{assumption}
\noindent
This mild assumption ensures that the function $f$ is bounded from below. We can now establish the following result.
\begin{theorem}
\label{Thm:Convergence}
Run PnP-ISTA with a denoiser~\eqref{Eq:MMSEDenoiser} under Assumptions~\ref{As:NonDeg}-\ref{As:GlobMin} using a fixed step-size $0 < \gamma \leq 1/L$. Then, the sequence $\{f(\xbm^t)\}_{t \geq 0}$ with $h$ defined in~\eqref{Eq:Regularizer} monotonically decreases and
$\|\nabla f(\xbm^t)\| \rightarrow 0$ as $t \rightarrow \infty$.
\end{theorem}
\noindent
The proof is provided in Appendix~\ref{Sec:ConvergenceAnalysis}. Theorem~\ref{Thm:Convergence} establishes convergence of PnP-ISTA with MMSE denoisers to a stationary point of the problem~\eqref{Eq:Optimization} where $h$ is specified in~\eqref{Eq:Regularizer}. The proof relies on the \emph{majorization-minimization (MM) strategy} widely used in the context of both convex and nonconvex optimization~\cite{Demptster.etal1977, Lange.etal2000, Beck.Teboulle2009, Razaviyayn.etal2016, Mairal2015}. It is important to note that the theorem does \emph{not} assume that $g$ or $h$ are convex, or that the denoiser is nonexpansive. The convexity of $\h$ is equivalent to the log-concavity of $p_\ybm$~\cite{Gribonval.Machart2013}, which does not hold for a wide variety of priors, such as mixtures of Gaussians~\cite{Teodoro.etal2019}. In fact, $\Dsf_\sigma$ is a proximal operator of a proper, closed, and convex function $h$ if and only if $\Dsf_\sigma$ is monotone and nonexpansive~\cite{Combettes.Pesquet2007}. Finally, note that $\h$ depends on both $\gamma$ and $\sigma$, both of which influence the relative weighting between $g$ and $h$. This is the consequence of $\h$ being specified by \emph{reverse engineering} the MMSE denoiser $\Dsf_\sigma$, which leads to the explicit dependence of the regularizer on the problem parameters~\cite{Gribonval2011}.

\begin{figure}[t]
        \begin{center}
                \includegraphics[width=8.5cm]{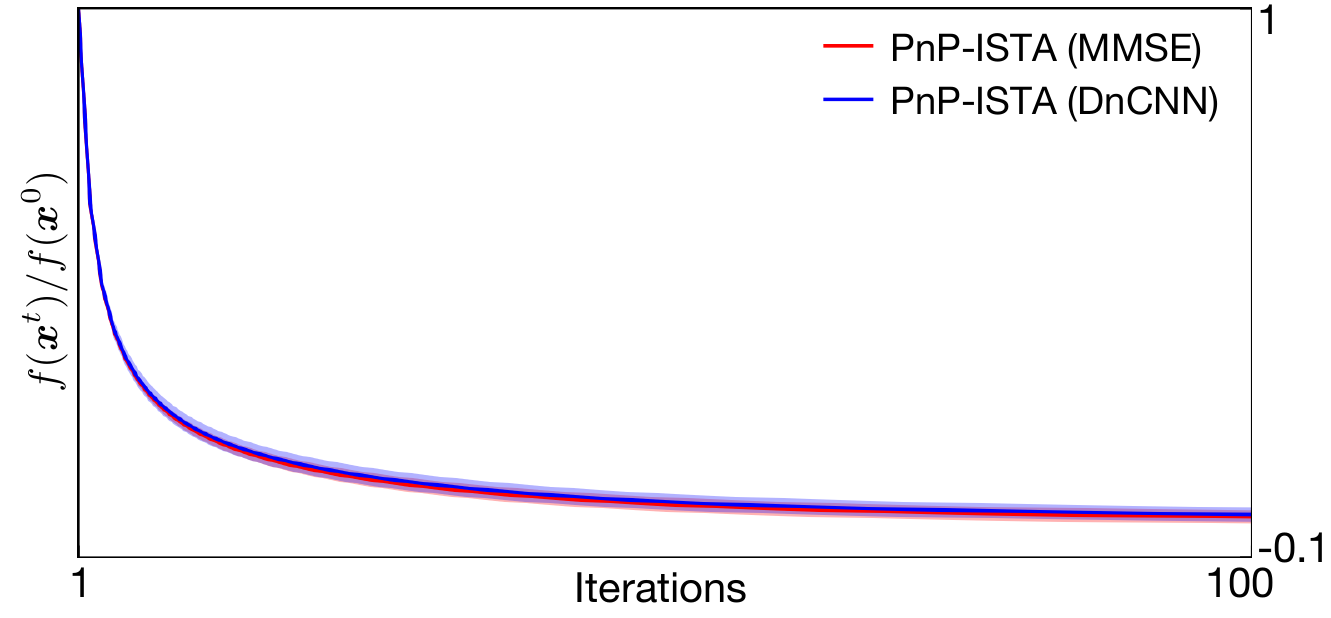}
        \end{center}
                \vspace{-0.75em}
        \caption{Convergence of PnP-ISTA for exact and approximate MMSE denoisers. The latter corresponds to DnCNN trained to minimize MSE. Average normalized cost $f(\xbm^t)/f(\xbm^0)$ is plotted against the iteration number with the shaded areas representing the range of values attained over 100 experiments. Note the monotonic decrease of the cost function $f$ as predicted by our analysis as well as the excellent agreement of both denoisers.}
        \label{Fig:ObjConv}
\end{figure}

\begin{figure}[t]
        \begin{center}
                \includegraphics[width=8.5cm]{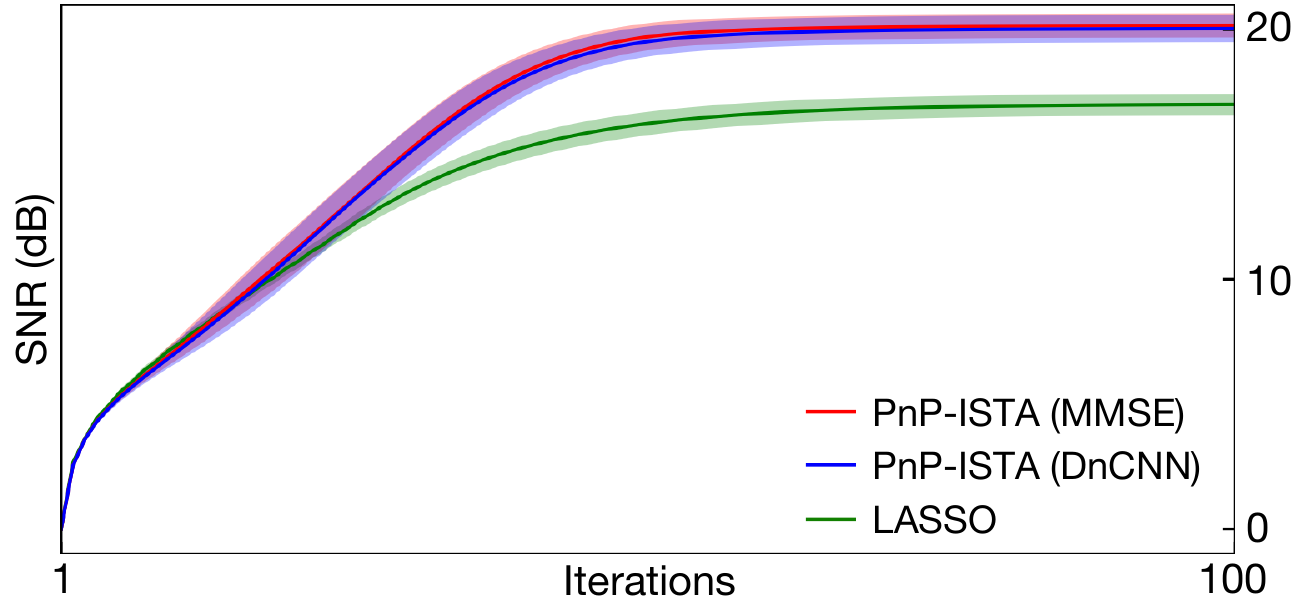}
        \end{center}
        \vspace{-0.75em}
        \caption{Convergence of PnP-ISTA for exact and approximate MMSE denoisers. The latter corresponds to DnCNN trained to minimize MSE. Average SNR (dB) is plotted against the iteration number with the shaded areas representing the range of values attained over 100 experiments. The SNR behavior of LASSO, implemented using ISTA with the $\ell_1$-norm prior, is also provided for reference. We highlight the excellent agreement of both denoisers and their superior SNR performance compared to the $\ell_1$ regularization.}
        \label{Fig:SNRConv}
\end{figure}

\begin{figure}[t]
        \begin{center}
                \includegraphics[width=8.5cm]{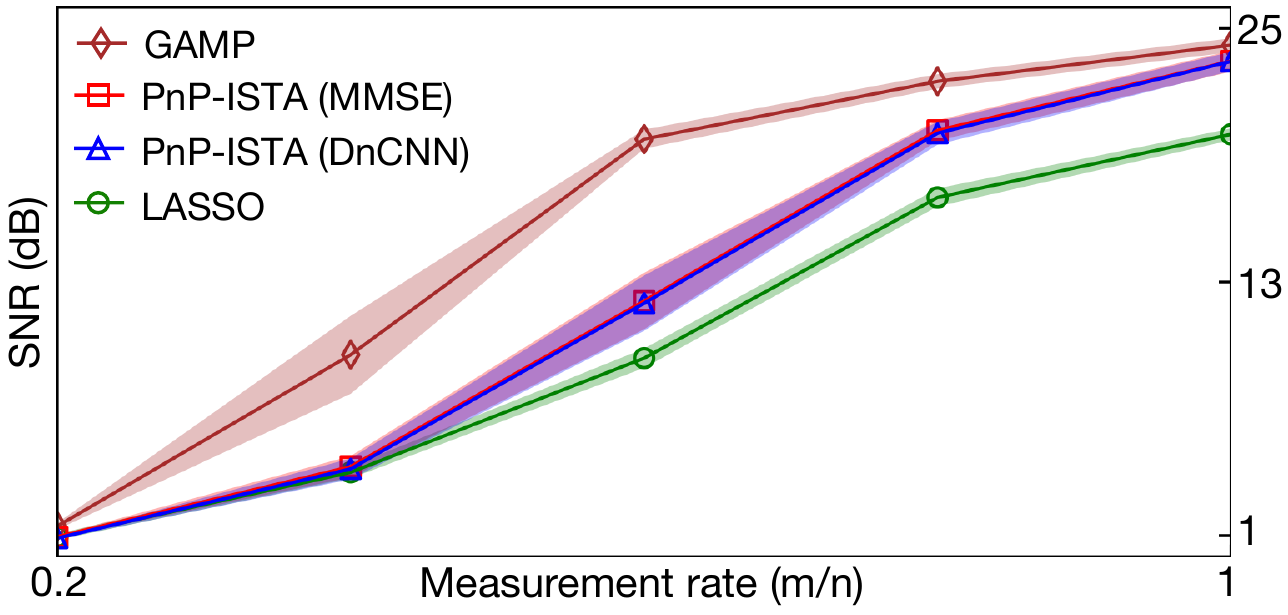}
        \end{center}
        \vspace{-0.75em}
        \caption{Illustration of the recovery performance of PnP-ISTA for exact and approximate MMSE denoisers. Average SNR (dB) is plotted against the measurement rate ($m/n$) with the shaded areas representing the range of values attained over 100 experiments. We also provide the performance of LASSO and GAMP, two widely used algorithms for sparse recovery in compressive sensing. The figure highlights the suboptimality of both variants of PnP-ISTA compared to GAMP, which stems from their assumption that errors in every ISTA iteration are AWGN. One can also observe the remarkable agreement between two variants of PnP-ISTA in all experiments.}
        \label{Fig:RecPerf}
\end{figure}


\section{Numerical Evaluation}

We illustrate PnP-ISTA with both \emph{exact} and \emph{approximate} MMSE denoisers on the problem of sparse signal recovery in compressive sensing~\cite{Candes.etal2006, Donoho2006}. We emphasize that our aim here is \emph{not} to argue that ISTA is a superior sparse recovery algorithm, or that the MMSE denoiser as a superior signal prior. Rather, we seek to gain new insights into the behavior of PnP-ISTA with MMSE priors in highly controlled setting.

As a model for the sparse vector $\xbm \in \R^n$ with $n =4096$, we consider a widely used independent and identically distributed (i.i.d.) Bernoulli-Gaussian distribution. Each component of $\xbm$ is thus generated from the distribution ${p_x(x) = \alpha \phi_{\sigma_x}(x) + (1-\alpha) \delta(x)}$, where $\delta$ is the Dirac delta function and
$\phi_{\sigma_x}$ is the Gaussian probability density function with zero mean and $\sigma_x > 0$ standard deviation. The parameter $0 \leq \alpha \leq 1$ in $p_x$ controls the sparsity of the signal, and we fix $\sigma_x^2 = 1/\alpha$. Since the distribution ${p_z = (\phi_\sigma \ast p_x)}$ is not log-concave~\cite{Gribonval2011}, the Bernoulli-Gaussian prior leads to a nonconvex regularizer and an expansive denoiser. The entries of $\Hbf \in \R^{m \times n}$ are generated as i.i.d.\ Gaussian random variables $\Ncal(0, 1/m)$. For each experiment, we additionally corrupt measurements with AWGN of variance $\sigma_e^2$ corresponding to an input SNR of 20 dB. Accordingly, the data fidelity term is set as least-squares $g(\xbm) = (1/2)\|\ybm - \Hbf \xbm \|^2$. All plots are obtained by averaging results over 100 random trials.

We consider two reference signal recovery algorithms extensively used in compressive sensing. The first is the standard \emph{least absolute shrinkage and selection operator (LASSO)}~\cite{Tibshirani1996}, which computes \eqref{Eq:Optimization} with an $\ell_1$-norm regularizer ${h(\xbm) = \lambda \|\xbm\|_1}$. The regularization parameter $\lambda > 0$ of LASSO is optimized for each experiment to maximize SNR. The second reference method is the MMSE variant of the \emph{generalized approximate message passing (GAMP)}~\cite{Rangan2011}, which is known to be nearly optimal for sparse signal recovery in compressive sensing~\cite{Krzakala.etal2012}. The parameters of GAMP are set to the actual statistical parameters $(\alpha, \sigma_x, \sigma_e)$ of the problem. While the suboptimality of ISTA to GAMP for random measurement matrices is well known, our aim is to illustrate the relative performance of ``optimal'' ISTA with the MMSE denoiser $\Dsf_\sigma$.

Since $\xbm$ is a vector with i.i.d.\ elements, the exact MMSE denoiser $\Dsf_\sigma$ can be evaluated as a sequence of scalar integrals. As an approximate MMSE denoiser, we use DnCNN with depth 4~\cite{Zhang.etal2017}. To that end, we train 9 different networks for the removal of AWGN at noise levels in the range from 0.01 to 0.37. The training was conducted over 2000 random realizations of the signal $\xbm \sim p_\xbm$ using the $\ell_2$-loss. For each experiment, we select the network achieving the highest SNR value under the scaling technique from~\cite{Xu.etal2020}.

Theorem~\ref{Thm:Convergence} establishes monotonic convergence of PnP-ISTA in terms of the cost function $f$. This is illustrated in Fig.~\ref{Fig:ObjConv} for the measurement rate $m/n = 0.8$. The average normalized cost $f(\xbm^t)/f(\xbm^0)$ is plotted against the iteration number for both exact and approximate MMSE denoisers. The shaded areas indicate the range of values taken over $100$ random trials. Fig.~\ref{Fig:SNRConv} illustrates the convergence behaviour of PnP-ISTA in terms of SNR (dB) for identical experimental setting by additionally including the SNR performance of LASSO as a reference. First, note the monotonic convergence of $\{f(\xbm^t)\}_{t \geq 0}$ as predicted by our analysis. Second, note the excellent agreement between two variants of PnP-ISTA. This close agreement is encouraging as deep neural nets have been extensively used as practical strategies for regularizing large-scale imaging problems.

The underlying assumption in PnP-ISTA is that errors within every ISTA iteration can be modeled as AWGN, which is known to be false~\cite{Donoho.etal2009}. This makes both exact and approximate MMSE denoisers ``suboptimal'' when used within PnP-ISTA. Unlike ISTA, GAMP explicitly ensures AWGN errors in every iteration \emph{for random measurement matrices}, making it a valid upper bound in our experimental setting. Fig.~\ref{Fig:SNRConv} illustrates the suboptimality of ``optimal'' ISTA for different measurement rates, highlighting the necessity of developing more accurate error models for PnP iterations~\cite{Eslahi.Foi2018}. Note again the remarkable agreement between DnCNN and the exact MMSE estimator, which highlights the practical relevance of our analysis.


\section{Conclusion}

This paper provides several new insights into the widely used PnP methodology by considering ``optimal'' MMSE denoisers. First, we have theoretically analyzed the convergence of PnP-ISTA for MMSE denoisers. Our analysis reveals that the algorithm converges even when the data-fidelity term is \emph{nonconvex} and denoiser is \emph{expansive}. This has not been shown in the prior work on PnP. Second, our simulations on sparse signal recovery illustrate the potential of \emph{approximate} MMSE denoisers---obtained by training deep neural nets---to match the performance of the \emph{exact} MMSE denoiser. The latter is intractable for high-dimensional imaging problems, while the former has been extensively used in practice. Third, our simulations highlight the \emph{suboptimality} of ``optimal'' ISTA with an MMSE denoiser, due to the assumption that error within ISTA iterations are Gaussian. We hypothesize that a similar phenomenon is present in the context of imaging inverse problems, which suggests the possibility of performance improvements by using more refined statistical models for characterizing errors within PnP algorithms~\cite{Eslahi.Foi2018}.


\appendix
\section{Appendix}
\label{Sec:MMSE_sigma}

\subsection{MMSE Denoising as Proximal Operator}
\label{Sec:MMSEDenoiser}

The relationship between MMSE estimation and regularized inversion has been established by Gribonval~\cite{Gribonval2011}, and has also been discussed in other contexts~\cite{Gribonval.Machart2013, Kazerouni.etal2013}. Our contribution is to formally connect this relationship to PnP algorithms, leading to their new interpretation for MMSE denoisers.

It is well known that the estimator~\eqref{Eq:MMSEDenoiser} can be compactly expressed using \emph{Tweedie's formula}~\cite{Robbins1956}
\begin{equation}
\label{Eq:Tweedie}
\Dsf_\sigma(\zbm) = \zbm - \sigma^2 \nabla h_\sigma(\zbm) \;\; \text{with} \;\; h_\sigma(\zbm) = -\log(p_\zbm(\zbm))\,,\!
\end{equation}
which can be obtained by differentiating~\eqref{Eq:MMSEDenoiser} using the expression for the probability distribution
\begin{equation}
\label{Eq:ConvRel}
p_\zbm(\zbm) = (p_\xbm \ast \phi_\sigma)(\zbm) = \int_{\R^n} \phi_\sigma(\zbm-\xbm)p_\xbm(\xbm) \d \xbm \;,
\end{equation}
where
$$\phi_\sigma(\xbm) \defn \frac{1}{(2\pi\sigma^2)^{\frac{n}{2}}} \exp\left(-\frac{\|\xbm\|^2}{2\sigma^2}\right) \;.$$
Since $\phi_\sigma$ is infinitely differentiable, so are $p_\zbm$ and $\Dsf_\sigma$. By differentiating $\Dsf_\sigma$, one can show that the Jacobian of $\Dsf_\sigma$ is positive definite (see Lemma 2 in~\cite{Gribonval2011})
\begin{equation}
\label{Eq:Jacobian}
\Jsf\Dsf_\sigma(\zbm) = \Ibf - \sigma^2 \Hsf h_\sigma(\zbm) \succ 0, \quad \zbm \in \R^n \;,
\end{equation}
where $\Hsf h_\sigma$ denotes the Hessian matrix of the function $h_\sigma$. Finally, Assumption~\ref{As:NonDeg} also implies that $\Dsf_\sigma$ is a \emph{one-to-one} mapping from $\R^n$ to $\Xcal = \Im(\Dsf_\sigma)$, which means that ${\Dsf^{-1}: \Xcal \rightarrow \R^n}$ is well defined and also infinitely differentiable over $\Xcal$ (see Lemma 1 in~\cite{Gribonval2011}). This directly implies that the regularizer $h$ in~\eqref{Eq:Regularizer} is also infinitely differentiable for any $\xbm \in \Xcal$.

We will now show that
\begin{align}
\label{Eq:DenoiserIsProx}
\Dsf_\sigma(\zbm) &= \prox_{\gamma h}(\zbm) \\
&= \argmin_{\xbm \in \R^n}\left\{\frac{1}{2}\|\xbm-\zbm\|^2 + \gamma h(\xbm)\right\}\nonumber
\end{align}
where $h$ is the (possibly nonconvex) function defined in~\eqref{Eq:Regularizer}. Our aim is to show that $\ubm^\ast = \zbm$ is the unique stationary point and global minimizer of
$$
\varphi(\ubm) \defn \frac{1}{2}\|\Dsf_\sigma(\ubm)-\zbm\|^2 + \gamma h(\Dsf_\sigma(\ubm))\,, \quad \ubm \in \R^n\;.
$$
By using the definition of $h$ in~\eqref{Eq:Regularizer} and the Tweedie's formula~\eqref{Eq:Tweedie}, we get
\begin{align*}
\varphi(\ubm)
&= \frac{1}{2}\|\Dsf_\sigma(\ubm)-\zbm\|^2 - \frac{1}{2}\|\Dsf_\sigma(\ubm)-\ubm\|^2 + \sigma^2 h_\sigma(\ubm) \\
&= \frac{1}{2}\|\Dsf_\sigma(\ubm)-\zbm\|^2 - \frac{\sigma^4}{2}\|\nabla h_\sigma(\ubm)\|^2 + \sigma^2h_\sigma(\ubm) \;.
\end{align*}
The gradient of $\varphi$ is then given by
\begin{align*}
&\nabla \varphi(\zbm) \\
&= [\Jsf\Dsf_\sigma(\ubm)](\Dsf_\sigma(\ubm)-\zbm) + \sigma^2 [\Ibf - \sigma^2 \Hsf h_\sigma(\ubm)]\nabla h_\sigma(\ubm) \\
&= [\Jsf \Dsf_\sigma(\ubm)]\left(\Dsf_\sigma(\ubm) + \sigma^2 \nabla h_\sigma(\ubm) - \zbm \right) \\
&= [\Jsf \Dsf_\sigma(\ubm)](\ubm-\zbm) \;,
\end{align*}
where we used~\eqref{Eq:Jacobian} in the second line and~\eqref{Eq:Tweedie} in the third line. Now consider a scalar function ${q(\nu) = \varphi(\zbm+\nu\ubm)}$ and its derivative
$$q'(\nu) = \nabla \varphi(\zbm+\nu\ubm)^\Tsf\ubm = \nu \ubm^\Tsf [\Jsf\Dsf_\sigma(\zbm+\nu\ubm)]\ubm \;.$$
From the positive definiteness of the Jacobian~\eqref{Eq:Jacobian}, we have $q'(\nu) < 0$ and $q'(\nu) > 0$ for $\nu < 0$ and $\nu > 0$, respectively. This implies that $\nu = 0$ is the global minimizer of $q$. Since $\ubm \in \R^n$ is an arbitrary vector, we have that $\varphi$ has no stationary point beyond $\ubm^\ast = \zbm$ and that $\varphi(\zbm) < \varphi(\ubm)$ for any $\ubm \neq \zbm$.

\subsection{Convergence Analysis}
\label{Sec:ConvergenceAnalysis}

Prior work has analyzed the convergence of PnP algorithms for contractive, nonexpansive, or bounded denoisers~\cite{Sreehari.etal2016, Chan.etal2017, Teodoro.etal2019, Sun.etal2019a, Ryu.etal2019, Gavaskar.Chaudhury2020}. Our analysis extends the prior work on PnP by analyzing convergence for MMSE denoisers without any assumptions on convexity of $g$ and $h$ or on nonexpansiveness of $\Dsf_\sigma$. We adopt the \emph{majorization-minimization (MM) strategy} widely used in nonconvex optimization~\cite{Demptster.etal1977, Lange.etal2000, Beck.Teboulle2009, Razaviyayn.etal2016, Mairal2015}.

Consider the following approximation of $f$ at $\sbm \in \R^n$
\begin{align*}
\mu(\xbm, \sbm)
&= g(\sbm) + \nabla g(\sbm)^\Tsf(\xbm-\sbm) + \frac{1}{2\gamma}\|\xbm-\sbm\|^2 + h(\xbm)\\
&= \frac{1}{2\gamma}\|\xbm-(\sbm-\gamma \nabla g(\sbm)\|^2 + h(\xbm) - \frac{\gamma}{2}\|\nabla g(\sbm)\|^2 \;.
\end{align*}
Assumption~\ref{As:Smooth} implies that for any $0 < \gamma \leq 1/L$, we have
\begin{equation}
\label{Eq:MajProp}
\mu(\xbm, \sbm) \geq f(\xbm) \quad\text{and}\quad \mu(\sbm, \sbm) = f(\sbm), \quad \xbm\,, \sbm \in \R^n \;.
\end{equation}
We express~\eqref{alg:pnpista} in the MM format
\begin{equation}
\label{Eq:MajMin}
\xbm^t = \argmin_{\xbm \in \R^n} \mu(\xbm, \xbm^{t-1}) = \Dsf_\sigma(\xbm^{t-1}-\gamma \nabla g(\xbm^{t-1})) \,,
\end{equation}
where from Appendix~\ref{Sec:MMSEDenoiser}, we know that $\Dsf_\sigma = \prox_{\gamma h}$.
Therefore, from~\eqref{Eq:MajProp} and~\eqref{Eq:MajMin}, we directly have that
$$f(\xbm^t) \leq \mu(\xbm^t, \xbm^{t-1}) \leq \mu(\xbm^{t-1}, \xbm^{t-1}) = f(\xbm^{t-1}) \;.$$
From Assumption~\ref{As:GlobMin}, we know that $f$ is bounded from below; therefore, the \emph{monotone convergence theorem} implies that the sequence $\{f(\xbm^t)\}_{t \geq 0}$ converges.

Consider the residual function $r$ between $\mu$ and $f$
\begin{align*}
r(\xbm)
&= \mu(\xbm, \sbm) - f(\xbm) \\
&= g(\sbm) + \nabla g(\sbm)^\Tsf(\xbm-\sbm) + \frac{1}{2\gamma}\|\xbm-\sbm\|^2-g(\xbm) \;.
\end{align*}
The definition of $r$ implies that $r(\sbm) = 0$ and $\nabla r(\sbm) = \zerobm$. Additionally, we have for any $\xbm, \ybm \in \R^n$
\begin{align*}
\|\nabla r(\xbm)-\nabla r(\ybm)\|
&= \|(1/\gamma)(\xbm-\ybm)-(\nabla g(\xbm)-\nabla g(\ybm))\| \\
&\leq (1/\gamma)\|\xbm-\ybm\| + \|\nabla g(\xbm)-\nabla g(\ybm)\| \\
&\leq (1/\gamma + L)\|\xbm-\ybm\| \leq (2/\gamma)\|\xbm-\ybm\| \;,
\end{align*}
where we used $0 < \gamma \leq 1/L$. The last inequality implies that $\nabla r$ is Lipschitz continuous with constant $2/\gamma$.

Denote by $f^\ast$ the infimum of $f$ and by
$$r_t(\xbm) = \mu(\xbm, \xbm^{t-1}) - f(\xbm) \geq 0, \quad \xbm \in \R^n \,,$$
the residual at iteration $t \geq 1$. Then,
\begin{align*}
&r_t(\xbm^t) = \mu(\xbm^t, \xbm^{t-1}) - f(\xbm^t) \leq f(\xbm^{t-1})-f(\xbm^t)\\
&\Rightarrow\quad \sum_{t = 1}^\infty r_t(\xbm^t) \leq (f(\xbm^0) - f^\ast) \;,
\end{align*}
where we used the fact that $f^\ast \leq \lim_{t\rightarrow \infty} f(\xbm^t)$.
This implies that $r_t(\xbm^t) \rightarrow 0$ as $t \rightarrow \infty$.

Since $\nabla r_t$ is $(2/\gamma)$-Lipschitz continuous, we have that
\begin{align*}
&\ubm \defn \xbm^t - \frac{\gamma}{2} \nabla r_t(\xbm^t) \\
&\Rightarrow\quad r_t(\ubm) \leq r_t(\xbm^t) - \frac{\gamma}{4}\|\nabla r_t(\xbm^t)\|^2 \;.
\end{align*}
Since $r_t(\xbm) \geq 0$, for all $\xbm \in \R^n$, we have
$$\|\nabla r_t(\xbm^t)\|^2 \leq \frac{4}{\gamma} (r_t(\xbm^t)-r_t(\ubm)) \leq \frac{4}{\gamma} r_t(\xbm^t) \rightarrow 0\;,$$
as $t \rightarrow \infty$.

Finally, consider the gradient of $f$ at $\xbm^t \in \Xcal = \Im(\Dsf_\sigma)$
$$\|\nabla f(\xbm^t)\| = \|\nabla_\xbm \mu(\xbm^t, \xbm^{t-1}) - \nabla r_t(\xbm^t)\| = \|r_t(\xbm^t)\| \rightarrow 0 \;,$$
as $t \rightarrow \infty$, where we used the fact that $\xbm^t$ is the minimizer of $\mu(\xbm, \xbm^{t-1})$. This concludes the proof.

\bibliographystyle{IEEEtran}

\end{document}